\documentclass{article}
\usepackage[square,sort,comma,numbers]{natbib}
\usepackage{arxiv}
\usepackage{natbib}
\pdfoutput=1
\usepackage{graphicx} 
\usepackage{amsmath}
\usepackage{amssymb}
\usepackage{solid_mechanics}
\usepackage{todonotes}
\usepackage{bm}
\usepackage{subcaption}
\usepackage{booktabs}
\title{Automated model discovery of finite strain elastoplasticity from uniaxial experiments}
\author{Asghar Jadoon\\
The University of Texas at Austin\\
Texas, USA \\
\And
Knut Andreas Meyer \\
TU Braunschweig \\
Germany\\
\And
Jan Fuhg \\
The University of Texas at Austin\\
Texas, USA 
}
\date{August 2024}

\newcommand{\dispmap}{\varphi}

\begin{document}

\maketitle

\section*{Abstract}
Constitutive modeling lies at the core of mechanics, allowing us to map strains onto stresses for a material in a given mechanical setting. Historically, researchers relied on phenomenological modeling where simple mathematical relationships were derived through experimentation and curve fitting. 
Recently, to automate the constitutive modeling process, data-driven approaches based on neural networks have been explored. While initial naive approaches violated established mechanical principles,  recent efforts concentrate on designing neural network architectures that incorporate physics and mechanistic assumptions into machine-learning-based constitutive models. For history-dependent materials, these models have so far predominantly been restricted to small-strain formulations. In this work, we develop a finite strain plasticity formulation based on thermodynamic potentials to model mixed isotropic and kinematic hardening. We then leverage physics-augmented neural networks to automate the discovery of thermodynamically consistent constitutive models of finite strain elastoplasticity from uniaxial experiments. We apply the framework to both synthetic and experimental data, demonstrating its ability to capture complex material behavior under cyclic uniaxial loading. Furthermore, we show that the neural network enhanced model trains easier than traditional phenomenological models as it is less sensitive to varying initial seeds. our model's ability to generalize beyond the training set underscores its robustness and predictive power. By automating the discovery of hardening models, our approach eliminates user bias and ensures that the resulting constitutive model complies with thermodynamic principles, thus offering a more systematic and physics-informed framework.

\section{Introduction}
One can gauge the significance of constitutive modeling by realizing that it has remained an active area of research over a long period of time to accurately model structural behavior. Historically, constitutive modeling aimed to provide an analytical expression that could map the strains in a material to the stresses that they induce. Numerous functional forms have been proposed to describe constitutive models for various material families including elastic and inelastic materials \cite{holzapfel2002nonlinear}. For any given material, one can simply obtain the parameters in these functional forms through laboratory testing and use the resulting expression to effectively model the material's constitutive behavior \cite{Simo_Hughes_2000, truesdell2004non}. 

With the recent advancements in computing power, machine learning has found applications in almost every discipline, including continuum mechanics \cite{10.3389/fmats.2019.00110}. Constitutive modeling in particular gained significant interest \cite{furukawa1998implicit, lefik2003artificial, jung2006neural}. The notion of using data-driven regression techniques for constitutive modeling is intuitive as both essentially map inputs to outputs. This has led to two primary streams of research. This first involves model-free approaches, which avoid empirical constitutive relations and directly leverage experimental data and therefore eliminate the traditional constitutive models \cite{KIRCHDOERFER201681, eggersmann2019model, kirchdoerfer2018data, ibanez2017data}. The second stream employs NNs to learn parametric constitutive relationships and substitutes them in place of traditional phenomenological constitutive laws in numerical solvers\cite{peirlinck2024automated, haghighat2023constitutive, flaschel2023automated}. Due to the need for large datasets\cite{Flaschel2023} for the model-free approach, this study focuses on the latter approach.

Following \cite{fuhg2024review} we distinguish between two approaches for NN-based constitutive modeling: (i) non-interpretable and (ii) interpretable. Non-interpretable approaches follow black-box learning routines and typically involve a large number of parameters whose influence cannot be construed in the physical sense. In contrast, interpretable approaches aim to find a functional form similar to traditional constitutive models, thereby enabling the definition of parameters and their effects on the physical system. Techniques such as symbolic and sparse regression \cite{flaschel2021unsupervised, 2310.03652, bomarito2021development,fuhg2024extreme}, have successfully been employed in this regard. However, these interpretable models are dependent on user-chosen functional form libraries which can lead to inferior approximation capabilities compared to non-interpretable approaches if chosen inappropriately \cite{peirlinck2024automated}. Prioritizing approximation accuracy over interpretability of the functional form, we adopt a non-interpretable but physically consistent approach in this study.
One of the pioneering efforts in this domain were Physics-Informed Neural Networks (PINNs) \cite{Karniadakis2021} which incorporate physical constraints weakly into the loss function. Another approach, recently termed Physics Augmented Neural Networks (PANNs)\cite{linden2023neural, klein2022finite}, enforces physical principles in a fundamentally different manner than PINNs. These networks are constrained to have a certain form of output with respect to the inputs and are designed to inherently satisfy all physical constraints by construction. The architectural details of these networks will be presented in subsequent sections.

PANNs have already been employed to good effect for various material models such as hyperelasticity, viscoelasticity, plasticity, and thermoelasticity \cite{flaschel2022discovering, wang2023automated, linka2023automated, tacc2023data,rosenkranz2024viscoelasticty,fuhg2024polyconvexneuralnetworkmodels}. However, for history-dependent materials, these models have so far predominantly been restricted to small-strain formulations\cite{meyer2023thermodynamically, fuhg2023modular} even though many engineering problems exhibit finite strain elastoplasticity. There have been extensive research efforts in the field of finite strain elastoplasticity since the last century leading to numerous monographs on the topic (see e.g. \cite{haupt2013continuum, lubliner2008plasticity, khan1995continuum, lemaitre1994mechanics} and some seminal works on the numerical treatment of these problems \cite{simo1998numerical, SIMO19881}. These principles have also been extended to anisotropic materials and coupled with damage mechanics \cite{ekh2004model, EKH20019461, vladimirov2010anisotropic}. However, this area of research has yet to utilize the mapping prowess of neural networks to effectively map strains onto stresses in a thermodynamically consistent manner even for relatively straightforward cases.

In this work, we develop a finite strain plasticity formulation based on thermodynamic potentials\cite{fuhg2023modular} motivated by previous studies\cite{Meyer2017,Meyer2018a,Meyer2021a} to model mixed isotropic and kinematic hardening. We then employ a physics-augmented neural network that satisfies all the thermodynamic constraints to learn and automate the discovery of such hardening models. This automated model-discovery framework eliminates user bias and ensures that the resulting constitutive model complies with thermodynamic principles, thus offering a more systematic and physics-informed framework. The paper is organized as follows: In Section 2, we present the theoretical framework for finite strain plasticity formulation based on thermodynamic potentials. In Section 3, we discuss various constraints on these potentials and the methods to enforce them in the neural network. The framework is tested against both synthetic and experimental data in Section 4 and Section 5 offers some concluding remarks.

\section{Finite strain plasticity formulation} \label{Theoretical formulation}
For our framework, we consider the current coordinates, $\tv{x}$, which can be obtained via the mapping, $\tv{x}=\dispmap(\tv{X},t)$, where $\tv{X}$ are the initial coordinates and $t$ the time. For the material modeling, we then assume a multiplicative decomposition (see e.g. \cite{haupt2013continuum}) of the deformation gradient, $\ts{F}:=\partial\dispmap/\partial\tv{X}$, into elastic, $\ts{F}\el$, and plastic, $\ts{F}\pl$, parts; $\ts{F}=\ts{F}\el\ts{F}\pl$. In addition to the plastic deformation gradient, we introduce the kinematic deformation gradient, $\ts{F}\kin$, and a scalar, k, as state variables to model kinematic and isotropic hardening, respectively. We introduce the following deformation tensors
\begin{subequations}
\begin{align}
    \ts{C}\el &= \tst{F}\el\ts{F}\el, \\
    \ts{c}\kin &= \tsti{F}\kin\tsi{F}\kin  \label{ckin}
\end{align}
\end{subequations}
as well as the velocity gradients 
\begin{subequations}
\label{eq:velocity_gradients}
\begin{align}
    \ts{L}\pl &:= \tsd{F}\pl\tsi{F}\pl, \\
    \ts{L}\kin &:= \tsd{F}\kin\tsi{F}\kin.
\end{align}
\end{subequations}

The Helmholtz free energy, $\varPsi$, is then defined in terms of the deformation tensors, as well as the internal variable $k$, in an additive fashion,
\begin{align}
    \varPsi := \varPsi\el(\ts{C}\el) + \varPsi\kin(\ts{c}\kin) + \varPsi\iso(k)
\end{align}
ensuring invariance wrt. rigid body rotations. 

We require that the material model fulfills the dissipation inequality,
\begin{align}
    \mathcal{D} :=& \ts{\tau}:\ts{l} - \dot{\varPsi} \geq 0 \nonumber \\
    =& \ts{\tau}:\ts{l} - \pdiff[\varPsi]{\ts{C}\el}:\tsd{C}\el - \pdiff[\varPsi]{\ts{c}\kin}:\tsd{c}\kin - \pdiff[\varPsi]{k} \dot{k}.
\end{align}
The Kirchhoff stress, $\ts{\tau}$, is given from the Cauchy stress, $\sig$, as $\ts{\tau}=J\sig$ where $J=\det(\ts{F})$ and the velocity gradient, $\ts{l}$, is defined as $\ts{l}:=\tvd{x}\otimes\tv{\nabla} = \tsd{F}\tsi{F}$. Following the standard Coleman–Noll procedure \cite{Coleman1963, holzapfel2002nonlinear}, and inserting the velocity gradients from Equation \eqref{eq:velocity_gradients}, we obtain the reduced dissipation inequality
\begin{align}
    \mathcal{D} = 2\ts{C}\el \pdiff[\varPsi]{\ts{C}\el}:\ts{L}\pl + 2\ts{c}\kin \pdiff[\varPsi]{\ts{c}\kin}:\ts{L}\kin - \pdiff[\varPsi]{k} \dot{k} \label{eq:initialdissipation}
\end{align}
along with the requirement
\begin{align}
    \ts{\tau} = 2\ts{F}\el\pdiff[\varPsi]{\ts{C}\el}\tst{F}\el.\label{eq:kirchhoff_stress}
\end{align}
We then introduce the Mandel stresses, $\ts{M}$ and $\ts{M}\kin$, and the isotropic hardening stress, $\kappa$,
\begin{subequations}
\begin{align}
    \ts{M} &:= 2\ts{C}\el \pdiff[\varPsi]{\ts{C}\el}, \\
    \ts{M}\kin &:= 2\ts{c}\kin \pdiff[\varPsi]{\ts{c}\kin}, \label{Mkin}\\
    \kappa &:= \pdiff[\varPsi]{k} \label{eq:kappa}
\end{align}
\end{subequations}
which following Equation \eqref{eq:kirchhoff_stress} yields 
\begin{align}
\ts{M}=\tst{F}\el \ts{\tau}\tsti{F}\el.  
\end{align}

Interpreting $\ts{M}\kin$ as a back-stress, we propose that the yield function, $\Phi$, depends on the difference, $\ts{M}-\ts{M}\kin$, as well as $\hat{\varPhi}\iso$ which is a function of the isotropic hardening stress, $\kappa$, i.e.
\begin{equation}
\Phi = f\subscr{y} (\bm{M} - \bm{M}_{\text{kin}}) - (Y_0 + \hat{\varPhi}\iso) \label{eq:yield}
\end{equation}

where $f\subscr{y}$ is the effective stress function and $Y_0$ the initial yield limit. This choice of yield function is novel in the sense that standard yield functions directly employ the isotropic hardening stress whereas we choose to replace it with the isotropic yield potential. Such a form of yield function allows us to introduce nonlinear isotropic hardening in a more straightforward manner. Thermodynamic constraints to ensure positivity of dissipation for such a formulation will be explained in the next section. 
For rate-independent plasticity, the Karush-Kuhn-Tucker conditions yield
\begin{align}
    \varPhi \leq 0, \quad \dot{\lambda} \geq 0, \quad \varPhi\dot{\lambda} = 0
\end{align}
with evolution laws for the internal variables as 

\begin{subequations}
\begin{align}
    \ts{L}\pl &:= \dot{\lambda} \pdiff[\varPhi\pl]{\ts{M}}, \label{Lp_evol} \\
    \ts{L}\kin &:= \dot{\lambda} \pdiff[\varPhi\kin]{\ts{M}\kin}, \label{Lkin_evol}\\
    \dot{k} &:= -\dot{\lambda} \pdiff[\varPhi\iso]{\kappa}.\label{k_evol}
\end{align}
\end{subequations}

\noindent
Equations \eqref{Lp_evol} and \eqref{Lkin_evol} are integrated using a standard exponential map \cite{korelc2014closed} whereas \eqref{k_evol} is integrated by the backward Euler rule to get

\begin{subequations}
\begin{align}
    \ts{F}\pl^{n+1} &= \exp \left ( \Delta \lambda  \pdiff[\varPhi\pl]{\ts{M}^{n+1}} \right) \ts{F}\pl^{n},\\
    \ts{F}\kin^{n+1} &= \exp \left( \Delta \lambda \pdiff[\varPhi\kin]{\ts{M}\kin^{n+1}} \right) \ts{F}\kin^{n}.\\
            k^{n+1} &=   -\Delta \lambda \pdiff[\varPhi\iso]{\kappa^{n+1}} +k^{n} .
\end{align}
\end{subequations}

In this work, we adopt the standard associative evolution of the plastic strains, $\varPhi\pl=\varPhi$, and further introduce the additive modification of the kinematic yield potential, $\varPhi\kin=\varPhi + \hat{\varPhi}\kin$ whereas the isotropic yield potential $\varPhi\iso=\varPhi $ is taken without the additive modification. As explained in the next section, such a form of isotropic yield potential enables us to model work hardening stagnation and represent saturation of the isotropic part, which is particularly useful when modeling the hysteresis loop. 
Doing so, the reduced dissipation becomes
\begin{align}
    \mathcal{D} &= \dot{\lambda} \left[ \left[\ts{M}-\ts{M}\kin\right]:\ts{\nu}  + \ts{M}\kin:\pdiff[\hat{\varPhi}\kin]{\ts{M}\kin} + \kappa \pdiff[\hat{\varPhi}\iso]{\kappa} \right], \quad \text{where} \quad \ts{\nu} := \pdiff[\varPhi]{\ts{M}} .\label{eq:finaldissipation}
\end{align}

\subsection{Standard model} \label{standardmodel}
We are now free to define the constitutive behavior by proposing expressions for the energy and dissipation potentials, as well as the effective stress. A strongly coupled behavior exists between the energy and dissipation potentials for both isotropic and kinematic hardening. The free energy potential $\varPsi\kin$ is a function of $\ts{c}\kin$ which, from Equation \eqref{ckin} depends on $\ts{F}\kin$. The update of $\ts{F}\kin$ during time integration comes from $\varPhi\kin$, or its derivative with respect to $\ts{M}\kin$ which, from Equation \eqref{Mkin}, depends on $\varPsi\kin$ or its derivative with respect to $\ts{c}\kin$. $\varPhi\iso$ and $\varPsi\iso$ also exhibit a similarly strong dependency on each other. Following the methodology in \cite{fuhg2023modular} and exploiting the modularity of our framework, we can choose which of these potentials we want to model using neural networks. Specifically, we explore three choices for modeling the potentials: (1) Utilize the standard functional form presented below, treating some of the constants or material parameters as trainable parameters within a gradient-based optimization framework; (2) Use the standard phenomenological forms for some potentials and use PANNs to represent the remaining potentials; or (3) Represent all potentials with thermodynamically consistent neural networks. In this study, we investigate and compare all of these approaches. As we assume that we only access to uniaxial experimental data, we, as described above, define the yield function as:
\begin{equation}
\Phi(\bm{M},\bm{M}_{\text{kin}}, Y_0, \kappa) = f_y (\bm{M} - \bm{M}_{\text{kin}}) - (Y_0 + \hat{\varPhi}\iso) \label{eq:yield}
\end{equation}
where $Y_0$ is the yield stress and $f_y$ represents the Von Mises effective stress defined as:
\begin{align}
    f\subscr{y}(\ts{M}\red) = \sqrt{\frac{3}{2} \ts{M}\red\dev:(\ts{M}\red^\mathrm{dev})^\mathrm{t}} \label{eq:vonmises}
\end{align}
where $\ts{M}\red\dev = \ts{M}\red - \ts{I} \tr(\ts{M}\red)/3$ and $\ts{M}\red=\ts{M}-\ts{M}\kin$. Next, we will present some standard choices for modeling the free energies and the dissipation potentials. The widely accepted Neo-Hookean formulation for compressible hyperelastic materials defines the elastic component of free energy as:
\begin{align}
    \varPsi\el(\ts{C}\el) &= \frac{G}{2}\left[\tr\left(\frac{\ts{C}\el}{\sqrt[3]{\det(\ts{C}\el)}}\right) - 3\right] + \frac{K}{2}\left[\sqrt{\det(\ts{C}\el)} - 1\right]^2 \label{NeoHookeanElasticFE}
\end{align}
where $G$ is the shear modulus, $K$ is the bulk modulus. Similar to the elastic part of the free energy, we have the following Neo-Hookean formulations for the isotropic and kinematic parts of free energy represented by the subscripts 'iso' and 'kin' respectively

\begin{subequations}

\begin{align}
    \varPsi\iso(k) &= \frac{H\iso}{2} k^2 ,\label{psiiso}\\
    \varPsi\kin(\ts{c}\kin) &= \frac{H\kin}{2}\left[\tr\left(\frac{\ts{c}\kin}{\sqrt[3]{\det(\ts{c}\kin)}}\right) - 3\right].
    \label{psikin}
\end{align}
    
\end{subequations}
Here $H\kin$ denotes the kinematic hardening modulus, and $H\iso$ is the isotropic hardening modulus. We can define the isotropic part of the dissipation potential using a nonlinear saturating potential as:
\begin{align}
    \hat{\varPhi}\iso &= R\subscr{sat,\infty} \left[1-\exp\left(-\gamma\: \kappa \right) \right] \label{voce_exp}
\end{align}

where $R\subscr{sat,\infty}$ represents the saturation stresses whereas parameter $\gamma$ controls the steepness of the curve or the degree of nonlinearity. Different forms of nonlinear kinematic (NLK) hardening exist (see \cite{Meyer2018a}). The simplest of them all is the Armstrong-Frederick kinematic hardening potential defined as:
\begin{align}
    \hat{\varPhi}\kin = \frac{3}{4} \frac{\ts{M}\kin\dev:(\ts{M}\kin^\mathrm{dev})^ \mathrm{t}}{M\subscr{kin,\infty}} \label{AFphikin}
\end{align}
where $M\subscr{kin,\infty}$ is the effective back-stress saturation value. A more complex model was proposed in \cite{OHNO1993375, OHNO1993391} called the Ohno-Wang model. In finite strains and in terms of Mandel stresses \cite{Meyer2018a} it relates the rate to
\begin{align}
    \pdiff[\hat{\varPhi}\kin]{\ts{M}\kin} = \frac{3}{2} \frac{\ts{M}\kin\dev}{M\subscr{kin,\infty}} \left(\frac{f_y (\bm{M} - \bm{M}_{\text{kin}})}{M\subscr{kin,\infty}}\right)^{m} \left\langle \frac{\ts{\nu}:\ts{M}\kin\dev}{f_y (\bm{M} - \bm{M}_{\text{kin}})} \right\rangle \label{OWphikin}
\end{align}
where $\langle \cdot \rangle$ are the Macaulay brackets inducing a different evolution of back-stresses for plastic loading and unloading whereas $m$ controls the rate of saturation.

\section{Finite strain plasticity with neural networks}
 We now work towards the main objective of this study i.e. to replace the energy and yield potentials with neural networks. Following \cite{meyer2023thermodynamically}, we are restricted by some mathematical and physical constraints that must be accounted for within the neural network framework. Although non-interpretable, enforcing these constraints gives us a reliable and physics-augmented framework. These constraints and their enforcement will now be examined in detail. The first constraint pertains to the convexity of free energy. Specifically, for nonlinear problems, the elastic component of the free energy must exhibit quasiconvexity \cite{Zee1983-ml} to guarantee material stability. Due to its nonlocality enforcing quasiconvexity in NNs is challenging, and therefore researchers have adopted the concept of polyconvexity (see e.g. \cite{klein2022polyconvex, fuhg2022learning, tacc2024benchmarking}) as outlined in \cite{Ball1976-as, ball1977constitutive}. Polyconvexity is a stricter condition that implies quasiconvexity and ensures material stability and the existence of a minimizer for variational functionals while being more straightforward to enforce in a neural network setting. Since NNs in (hyper)elastic problems have already been investigated\cite{kalina2022automated, tac2023benchmarks, chen2022polyconvex} extensively and we are concerned with single uniaxial loading experiments, we will not focus on the elastic component of the strain energy function in this work and rely on standard Neo-Hookean free energy from Equation \eqref{NeoHookeanElasticFE}. Such a form inherently fulfills the polyconvexity requirements. We remark however that in cases where more experimental loading data is available, one can easily replace this formulation with a PANN.
 
 We will instead focus on 
  neural network representations of the isotropic and kinematic hardening components. To this end, we will adopt the following nomenclature
\begin{itemize}
    \item Neural Network for $ \varPsi\kin(\ts{c}\kin)$ $\rightarrow$ $\mathcal{NN}_{\varPsi\kin}(\cdot)$
    \item Neural Network for $ \varPsi\iso(k)$ $\rightarrow$ $\mathcal{NN}_{\varPsi\iso}(k)$
    \item Neural Network for $ \hat{\varPhi}\kin(\ts{M}\kin)$ $\rightarrow$ $\mathcal{NN}_{\hat{\varPhi}\kin}(\cdot)$
    \item Neural Network for $ \hat{\varPhi}\iso(\kappa)$ $\rightarrow$ $\mathcal{NN}_{\hat{\varPhi}\iso}(\kappa)$
\end{itemize}
 \noindent Here, ($\cdot$) represents the choice of invariants to be used. As pointed out earlier, we require polyconvexity of the elastic part of free energy while the free energies associated with kinematic and isotropic hardening need to be convex. Therefore, we must ensure the convexity of $\mathcal{NN}_{\varPsi\kin}$ and $\mathcal{NN}_{\varPsi\iso}$.
 
 We are also constrained to have a model that is independent of the observer, ensuring frame invariance and objectivity. To satisfy this constraint, energy and dissipation potentials are usually expressed in terms of invariants. However, doing so necessitates additional consideration towards the convexity of these potentials. According to \cite{Boyd_Vandenberghe_2011}, a function $f(x)=h(g(x))$ is convex if both $g$ and $h$ are convex and $h$ is nondecreasing. Thus, the invariants must be convex functions of the input tensor and the neural network must be convex and monotonically nondecreasing to ensure convexity with respect to the tensor itself. Besides ensuring frame invariance, the use of invariants helps reduce the number of inputs to the neural networks while still containing maximum information about the tensors. Of course, this is irrelevant for the isotropic components of energy dissipation potentials since they depend only on a scalar. Accordingly, the kinematic part of the free energy $\varPsi\kin(\ts{c}\kin)$ will be expressed in terms of invariants of $\ts{c}\kin$. Since we assume access to single uniaxial loading cases we select the following isotropic invariants as input to $\mathcal{NN}_{\varPsi\kin}$:
\begin{equation}
    \bar{I_1} = \tr(\ts{c}\kin), \quad \bar{I_2} = \tr(\ts{c}\kin^2)
\end{equation}
Although these invariants don't encompass deviatoric information explicitly, due to the coupled nature of free energies and dissipation potentials, we can introduce the deviatoric information through the invariants of kinematic dissipation potential. We must also ensure the symmetry of stresses. While the Mandel stresses are symmetric under elastic isotropy, we focus on the back stresses. Given that $\varPsi\kin = \mathcal{NN}_{\varPsi\kin}(\bar{I_1},\bar{I_2})$, we have
\begin{equation}
    \begin{aligned}
           \ts{M}\kin :&= 2\ts{c}\kin \pdiff[\varPsi\kin]{\ts{c}\kin} = 2\pdiff[\mathcal{NN}_{\varPsi\kin}]{\bar{I_1}} \ts{c}\kin + 4\pdiff[\mathcal{NN}_{\varPsi\kin}]{\bar{I_2}} \ts{c}\kin \cdot \ts{c}\kin  
    \end{aligned}
\end{equation}
showing that $\ts{M}\kin$ is symmetric in our case. Thermodynamic consistency demands the non-negativity of dissipation. Although ensuring that the summation of all terms in Equation \eqref{eq:finaldissipation} remains positive is challenging, if we can ensure all the terms individually are positive, we can ensure non-negative dissipation and thermodynamic consistency. For brevity, we present only the constraints on potentials here but interested readers can refer to Appendix A for a detailed derivation of these constraints. $\mathcal{NN}_{\hat{\varPhi}\kin}$, now a function of invariants of $\ts{M}\kin$, has the following constraints:
\begin{itemize}
      \item Convexity with respect to all the invariants
      \item Monotonically non-decreasing with respect to all invariants
      \item $\mathcal{NN}_{\hat{\varPhi}\kin}(\ts{I}) = 0$
      \item $\mathcal{NN}_{\hat{\varPhi}\kin}(x) \geq 0$ for all $x$
\end{itemize}
The constraints on $\mathcal{NN}_{\hat{\varPhi}\kin}$ restrict us to the use of positive invariants since $\mathcal{NN}_{\hat{\varPhi}\kin}(\ts{I}) = 0$ and $\mathcal{NN}_{\hat{\varPhi}\kin}$ must be monotonically increasing, which implies:
\begin{equation}
\mathcal{NN}_{\hat{\varPhi}\kin}(\ts{I}) \geq \mathcal{NN}_{\hat{\varPhi}\kin}(\ts{x}) \quad \forall \ts{x} \in [-\infty, 0].
\end{equation}
Here, $\ts{I}$ represents the invariant values at the undeformed configuration. Coupled with the positivity constraint, the only way for the relation above to hold and still satisfy all the constraints is if:
\begin{equation}
\mathcal{NN}_{\hat{\varPhi}\kin}(\ts{x}) = 0 \quad \forall \ts{x} \in [-\infty, 0].
\end{equation}
Therefore, $\mathcal{NN}_{\hat{\varPhi}\kin}$ will always be zero for any negative inputs, and we must choose positive invariants instead. Therefore, we use the following invariants:
\begin{equation}
    \hat{I_1} = \ts{M}\kin : \ts{M}\kin, \quad \hat{I_2} = (\ts{M}\kin : \ts{\eta_v})^2  , \quad \ts{\eta_v} := \frac{\ts{\nu}}{\lVert \ts{\nu} \rVert}
\end{equation}
Since we are primarily interested in the derivatives of these potentials, this choice of invariants helps us get a similar form as the modified Burlet–Cailletaud  model\cite{burlet1986numerical} presented in \cite{Meyer2018a} where the first invariant gives the Armstrong-Frederick\cite{armstrong1966mathematical} type hardening in the evolution equation and the second invariant gives the radial evanescence term. Additionally, for $\mathcal{NN}_{\hat{\varPhi}\iso}$, we have the following constraints:
   \begin{itemize}
      \item Monotonically nondecreasing with respect to $\kappa$.
      \item $\mathcal{NN}_{\hat{\varPhi}\iso}(\kappa) \geq 0$ for all $\kappa$
      \item $\mathcal{NN}_{\hat{\varPhi}\iso}(0) = 0$
   \end{itemize}
Note that such a formulation is considerably less restrictive on $\hat{\Phi}_{\text{iso}}$ and allows us to model the saturation of stresses associated with isotropic hardening. The first two conditions stem from the fact that we are only considering isotropic hardening. If we wanted to model isotropic softening, $\kappa$ would be negative and that would require $\hat{\Phi}_{\text{iso}}$ to be nonincreasing. While that would fulfill the requirements for thermodynamic consistency, we would still have a positive $\hat{\Phi}_{\text{iso}}$ even for a negative $\kappa$. Therefore, this formulation can only model isotropic hardening.

We also have a convexity constraint on $\varPhi$ with respect to $\ts{M}$ so we can propose an expression for the effective stress $f\subscr{y}(\ts{M}\red)$ which must be convex and monotonically nondecreasing in terms of the invariants of $\ts{M}$ in $\pi_{1}$ and $\pi_{2}$, i.e. the $\pi$-plane components of $\ts{M}$ with $\pi_{3}=\text{constant}$ for pressure-insensitive materials. Within this framework, we use the Von-Mises effective stress from Equation \eqref{eq:vonmises} which satisfies these constraints.

 Several studies (see e.g. \cite{YOSHIDA2002661, MONTANS2000135, HE2019507, ZHONG2022107396}) employ a 'bounding surface' or a 'memory surface' for modeling NLK hardening. This two-surface approach is generally identical with a multi-NLK model i.e. a superposition of multiple NLK hardening components. Such an approach simply boils down to the superposition of multiple backstresses i.e.


\begin{equation}
\ts{M}\kin = \sum^{n}_{i=1} \ts{M}\kini
\end{equation}
where $n$ is the number of backstresses we want. Consequently, we get multiple deformation gradients for kinematic hardening from separate evolution laws:
\begin{equation}
    \ts{L}\kini = \dot{\lambda} \pdiff[\varPhi\kin]{\ts{M}\kini}
\end{equation}
While we could employ distinct neural networks to model the kinematic dissipation potential for each backstress, we opt for a single potential in this work. We can finally summarize the constraints on the neural networks:
\begin{itemize}
    \item $\mathcal{NN}_{\varPsi\kin}(\bar{I_1},\bar{I_2})$ $\rightarrow$ Convex and monotonically nondecreasing
    \item $\mathcal{NN}_{\varPsi\iso}(k)$ $\rightarrow$ Convex and monotonically nondecreasing
    \item $\mathcal{NN}_{\hat{\varPhi}\kin}(\hat{I_1},\hat{I_2})$$\rightarrow$  Positive, convex and monotonically nondecreasing
    \item $\mathcal{NN}_{\hat{\varPhi}\iso}(\kappa)$ $\rightarrow$ Positive and monotonically nondecreasing
\end{itemize}

   \begin{figure}[t]
    \centering
    \includegraphics[width=1.0\textwidth]{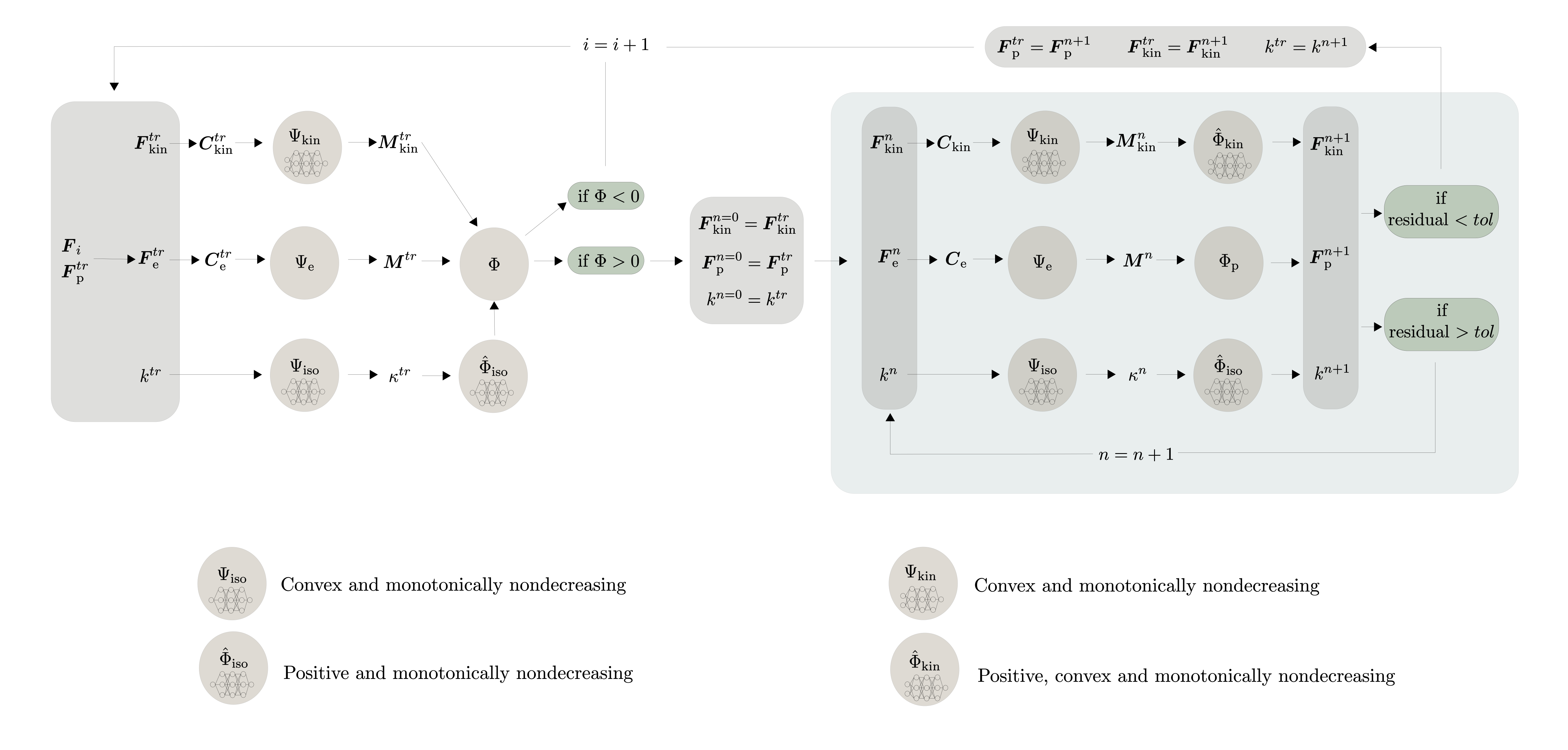}
    \caption{Simulation schematic for our analysis. The subscript $i$ represents the pseudo time step whereas subscript $n$ denotes the Newton-Raphson iteration for the evolution of internal variables which continues until  a specified tolerance $tol$ is achieved for the residual.}
    \label{fig:simulation}
\end{figure}

A simulation schematic is presented in Figure \ref{fig:simulation}. Now that we have established the constraints on our potentials, following \cite{amos2017input}, we can create input-convex neural networks and with some modifications of the neural network architecture, we can enforce all our constraints. For a neural network with layers $l=1,2\ldots,L$, weights $W$, bias $b$ and activation functions $\sigma$, we require the following architectures:
\begin{itemize}
    \item For a neural network to be \textbf{convex}, and \textbf{monotonically increasing}, we need:

\begin{itemize}
    \item Non-negative weights in all layers: 
    \[
    W \geq 0 \quad \forall \, l
    \]
    \item Positive, monotonically increasing and convex activation functions:
    \[
    \sigma(x) > 0 \quad \text{and} \quad \sigma'(x) \geq 0  \quad \text{and} \quad \sigma''(x) \geq 0
    \]
\end{itemize}
\end{itemize}

\begin{itemize}
    \item For a neural network to be \textbf{positive}, \textbf{convex}, and \textbf{monotonically increasing}, we need:

\begin{itemize}
    \item Non-negative weights in all layers: 
    \[
    W \geq 0 \quad \forall \, l
    \]
    \item Positive, monotonically increasing and convex activation functions:
    \[
    \sigma(x) > 0 \quad \text{and} \quad \sigma'(x) \geq 0  \quad \text{and} \quad \sigma''(x) \geq 0
    \]
    \item Positive bias in the last layer:
    \[
    b^{(L)} \geq 0
    \]
\end{itemize}
\end{itemize}

\begin{itemize}
    \item For a neural network to be \textbf{positive} and \textbf{monotonically increasing}, we need:

\begin{itemize}
    \item Non-negative weights in all layers: 
    \[
    W \geq 0 \quad \forall \, l
    \]
    \item Positive and monotonically increasing activation functions:
    \[
    \sigma(x) > 0 \quad \text{and} \quad \sigma'(x) \geq 0
    \]
    \item Positive bias in the last layer:
    \[
    b^{(L)} \geq 0
    \]
\end{itemize}
\end{itemize}

Following \cite{fuhg2023modular}, we define parametric activation functions for the neural networks. We select a parametric form of the Softplus and logistic activation functions for networks requiring convexity and monotonicity respectively as:
\begin{equation}
\mathcal{A}_c(x) = \log\left(1 + e^\alpha e^x\right) \quad \mathcal{A}_m(x) = \frac{\beta_1}{1 + e^{-\beta_2 (x - \beta_3)}} 
\end{equation}
Here, the subscripts $c$ and $m$ represent convex and monotonically nondecreasing respectively with $\mathcal{A}_c$ to be used for all the neural networks that require convexity and $\mathcal{A}_m$ used for $\mathcal{NN}_{\hat{\varPhi}\iso}(\kappa)$ which is only monotonically nondecreasing. Similar to $\alpha$, $\beta_3$ can take on any value as it defines translation along the x-axis, whereas $\beta_1$, $\beta_2$ need to be greater than zero to ensure a monotonically nondecreasing form. Furthermore, $\beta_1$ and $\beta_2$ control the supremum and steepness of the function respectively.
The training routine was implemented in a PyTorch \cite{paszke2017automatic} environment, relying on the Adam optimizer \cite{kingma2014adam} and a mean squared error defined as
\begin{align}
\mathcal{L} = \frac{1}{N} \sum_{i=1}^N [x_i - \hat{x}_i]^2
\end{align}
where $N$ is the batch size, $x_i$ represents the predicted output at time step $i$ and $\hat{x}_i$ is the actual response at a given time step.

\section{Results}
In this section, we present the results for three different modeling approaches discussed earlier, evaluated on both synthetic and experimental data. The first approach utilizes the standard forms of energy and dissipation potentials outlined in Section \ref{standardmodel} while training their parameters. Here, we explore modeling the kinematic dissipation potential either with Armstrong-Frederick (AF) or with an Ohno-Wang (OW) kinematic hardening model. In the second approach, termed '2NN', we fix the free energy potentials to the Neo-Hookean forms presented in Section \ref{standardmodel} while treating the material constants as trainable parameters and using PANNs for the dissipation potentials. The final approach referred to as '4NN', involves the use of PANNs for all energy and dissipation potentials. These chocies are summarized in table \ref{tab:PotentialChoices_Frameworks}. All approaches were tested with different initializations of the parameters to test their robustness against varying initial seeds. All neural networks were modeled with 2 hidden layers with 20 neurons in each layer. The stress responses for all the frameworks are plotted against the stretch $\lambda$ with $\sigma_{11}$ representing the components of Cauchy stress (GPa) along the direction of uniaxial compression or tension.

\begin{table}[h]
  \centering
  \caption{Potentials employed in each framework.}
  \label{tab:PotentialChoices_Frameworks}
  \begin{tabular}{ccccc}
    \toprule
    \textbf{Model} & \textbf{AF} & \textbf{OW} & \textbf{2NN} & \textbf{4NN}\\
    \midrule
    $\varPsi\iso$ & Eq. \eqref{psiiso} & Eq. \eqref{psiiso} & Eq. \eqref{psiiso} & $\mathcal{NN}_{\varPsi\iso}$\\
    $\varPsi\kin$ & Eq. \eqref{psikin} & Eq. \eqref{psikin} & Eq. \eqref{psikin} & $\mathcal{NN}_{\varPsi\kin}$\\
    $\hat{\varPhi}\iso$ & Eq. \eqref{voce_exp} & Eq. \eqref{voce_exp} & $\mathcal{NN}_{\hat{\varPhi}\iso}$ & $\mathcal{NN}_{\hat{\varPhi}\iso}$\\
        $\hat{\varPhi}\kin$ & Eq. \eqref{AFphikin} & Eq. \eqref{OWphikin} & $\mathcal{NN}_{\hat{\varPhi}\kin}$ & $\mathcal{NN}_{\hat{\varPhi}\kin}$\\
    \bottomrule
  \end{tabular}
\end{table}

\subsection{Synthetic Data}
\begin{figure}
     \centering
     \begin{subfigure}{0.4\textwidth}
         \centering
         \includegraphics[width=\textwidth]{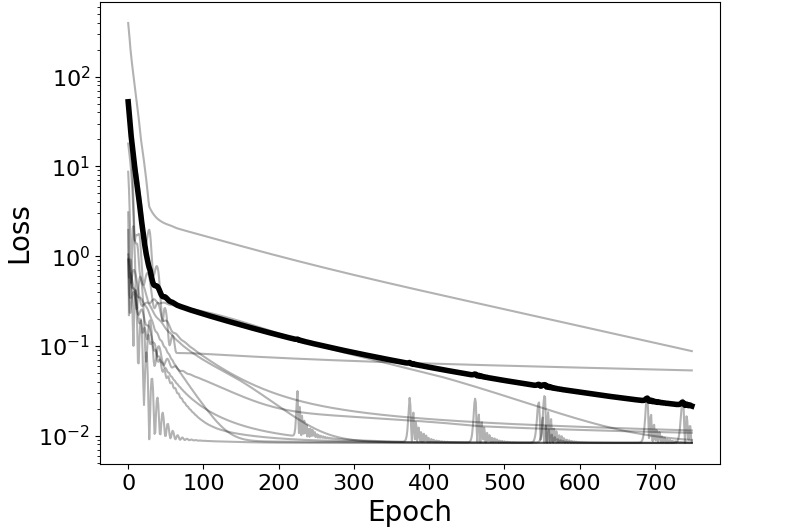}
         \caption{}
         \label{OWLossBC2}
     \end{subfigure}
     \hspace{0.05\textwidth} 
     \begin{subfigure}{0.4\textwidth}
         \centering
         \includegraphics[width=\textwidth]{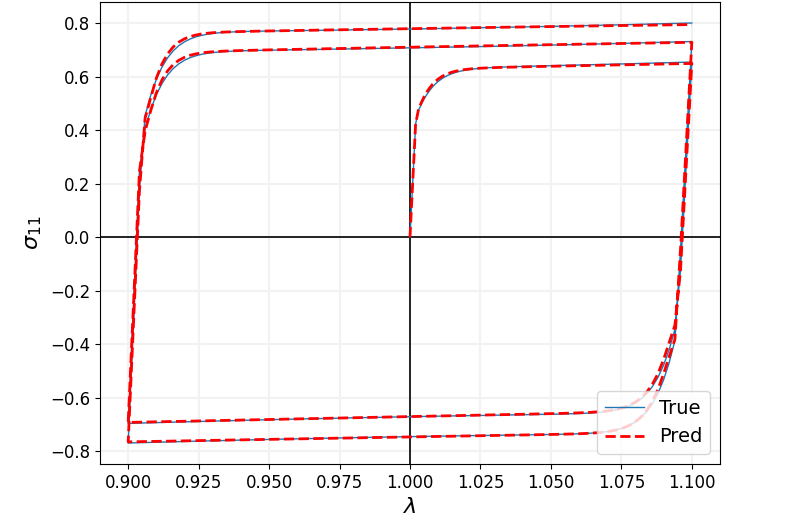}
         \caption{}
         \label{OWResponseBC2}
     \end{subfigure}
     
     \caption{Results for OW on the first synthetic dataset. (a) Evolution of the loss during the training routine for fitting the parameters to Ohno-Wang model for kinematic hardening. The lighter lines represent the loss evolution for different initialization whereas the darker line represents the mean loss. (b) True and predicted stress values from the best loss response.}
     \label{OWDataBC2}
\end{figure}

\begin{figure}
     \centering
     \begin{subfigure}{0.4\textwidth}
         \centering
         \includegraphics[width=\textwidth]{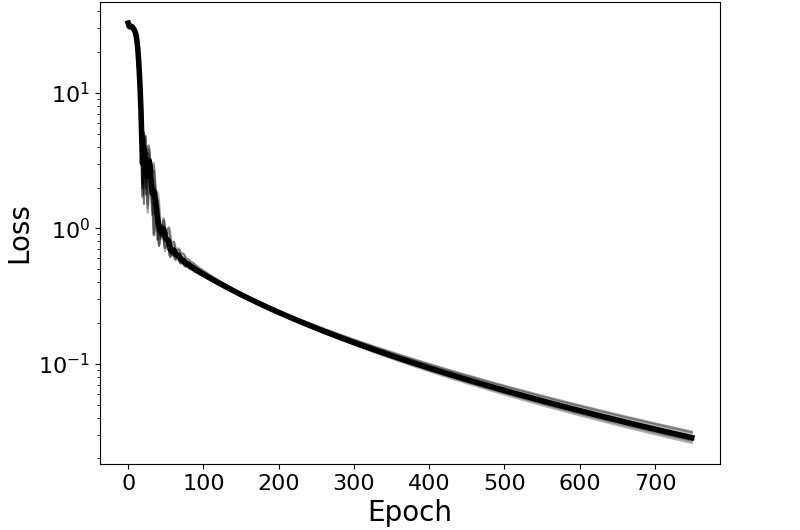}
         \caption{}
         \label{2NNLossBC2}
     \end{subfigure}
     \hspace{0.05\textwidth} 
     \begin{subfigure}{0.4\textwidth}
         \centering
         \includegraphics[width=\textwidth]{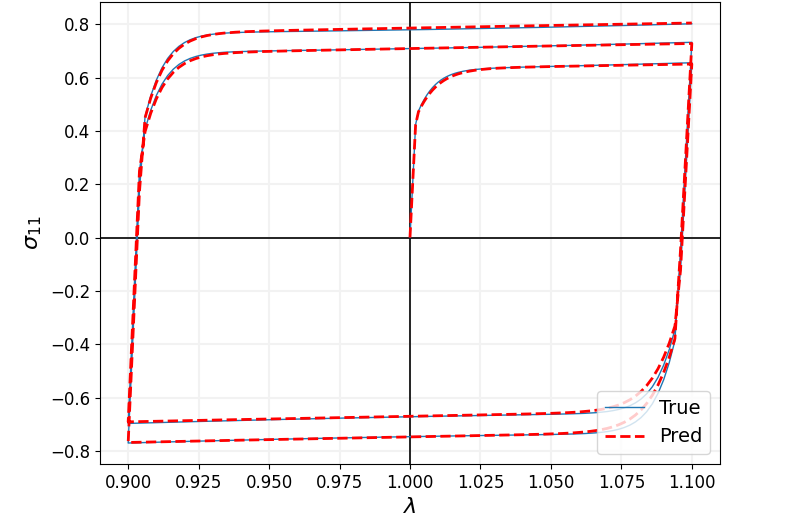}
         \caption{}
         \label{2NNResponseBC2}
     \end{subfigure}
          \begin{subfigure}{0.4\textwidth}
         \centering
         \includegraphics[width=\textwidth]{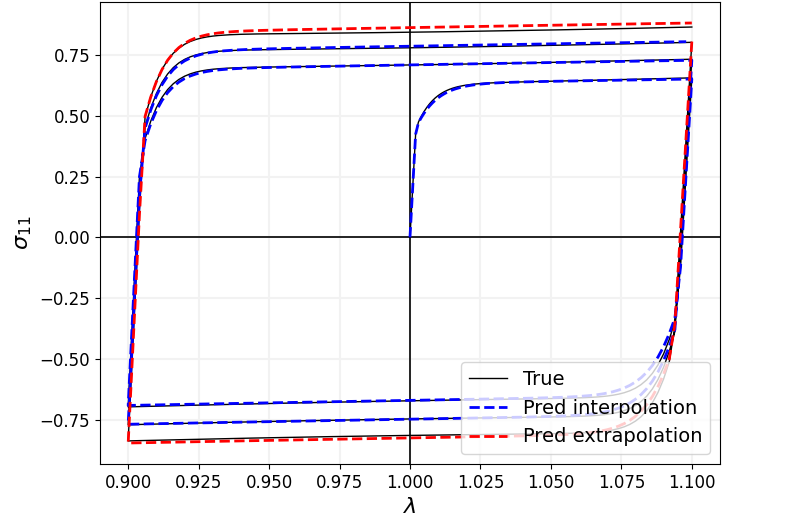}
         \caption{}
         \label{2NNExtrapolatedBC2}
     \end{subfigure}
     
     \caption{Results for 2NN on the first synthetic dataset. (a) Evolution of the loss during the training routine for PANNs for dissipation potentials and fitting the parameters of Neo-Hookean free energies. The lighter lines represent the loss evolution for different initialization whereas the darker line represents the mean loss. (b) True and predicted stress values from the best loss response. (c) Model's performance on unseen data.}
     \label{2NNDataSyn}
\end{figure}

\begin{figure}
     \centering
     \begin{subfigure}{0.4\textwidth}
         \centering
         \includegraphics[width=\textwidth]{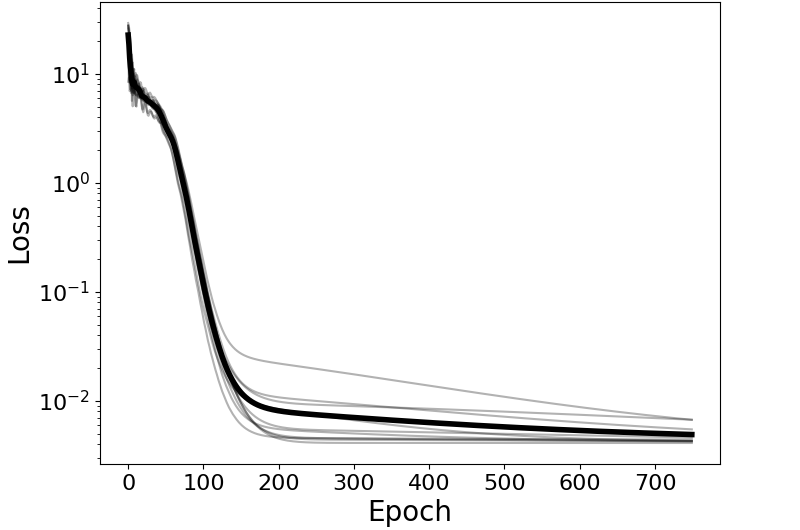}
         \caption{}
         \label{4NNLossBC2}
     \end{subfigure}
     \hspace{0.05\textwidth} 
     \begin{subfigure}{0.4\textwidth}
         \centering
         \includegraphics[width=\textwidth]{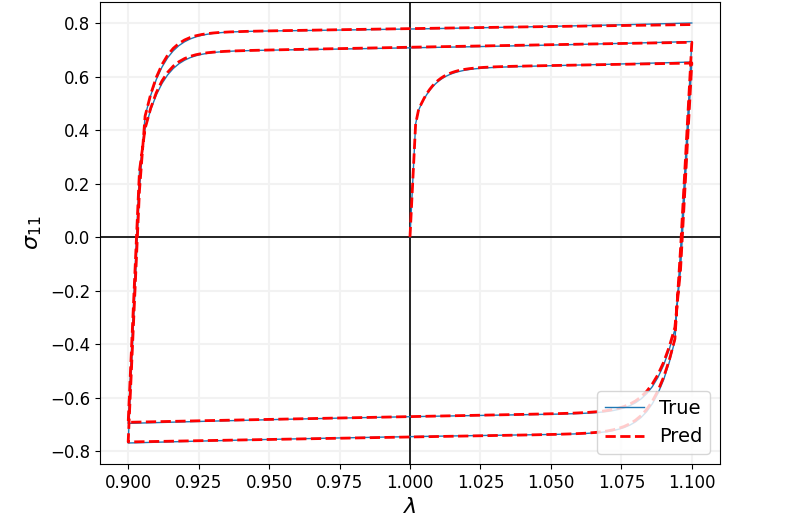}
         \caption{}
         \label{4NNResponseBC2}
     \end{subfigure}
          \begin{subfigure}{0.4\textwidth}
         \centering
         \includegraphics[width=\textwidth]{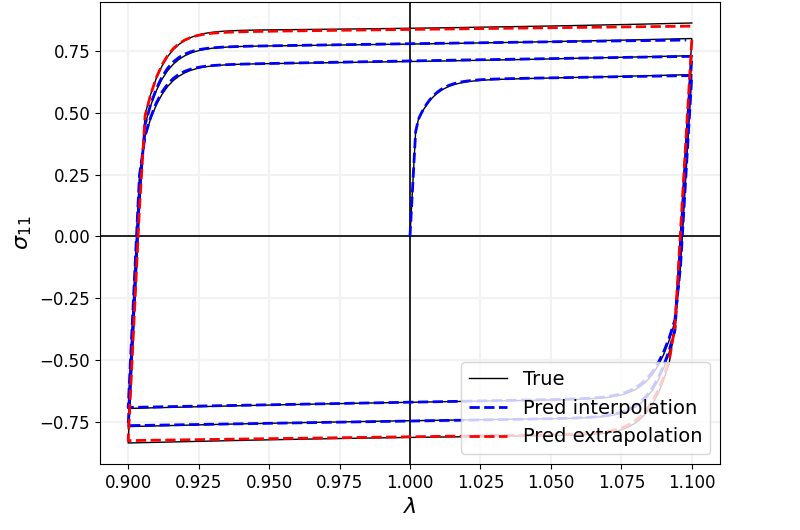}
         \caption{}
         \label{4NNExtrapolationBC2}
     \end{subfigure}
     
     \caption{Results for 4NN on the first synthetic dataset. (a) Evolution of the loss during the training routine for PANNs for all the potentials. The lighter lines represent the loss evolution for different initialization whereas the darker line represents the mean loss. (b) True and predicted stress values from the best loss response. (c) Model's performance on unseen data.}
     \label{4NNDataSyn}
\end{figure}

We first employ our framework on two synthetic datasets generated for cyclic uniaxial loading. The first synthetic dataset was produced using the modified Burlet–Cailletaud model \cite{burlet1986numerical} presented in \cite{Meyer2018a} to represent multi-NLK hardening. Since the Burlet–Cailletaud model is equivalent to the Armstrong-Frederick model under uniaxial loading, we only test this dataset with the Ohno-Wang model for the parameter-fitting case. Figure \ref{OWDataBC2} shows the evolution of training losses and the best stress fits achieved while training the parameters of the Ohno-Wang kinematic hardening model for the first dataset. Figures \ref{2NNLossBC2} and \ref{2NNResponseBC2} show the same for the 2NN approach, where parameters of Neo-Hookean free energies were trained and PANNs were used for dissipation potentials. Figures \ref{4NNLossBC2} and \ref{4NNResponseBC2} show the loss evolution and best stress fit for the 4NN approach, utilizing PANNs for all energy and dissipation potentials. Extrapolation capabilities of the PANN-based approaches were also tested as shown in Figures \ref{2NNExtrapolatedBC2} and \ref{4NNExtrapolationBC2}. The model was trained on 5 cycles of uniaxial loading and was then subjected to an additional two cycles for the model to extrapolate. As mentioned earlier, all frameworks were tested with 10 different parameter initializations and Table \ref{tab:SynData_loss_table_BC2} compares the approaches based on the lowest loss attained, mean loss across all runs, and the standard deviation.

\begin{table}
  \centering
  \caption{Loss for each framework for synthetic data generated with modified Burlet–Cailletaud model \cite{Meyer2018a}.}
  \label{tab:SynData_loss_table_BC2}
  \begin{tabular}{cccc}
    \toprule
    \textbf{Model} & \textbf{Lowest Loss} & \textbf{Mean Loss} & \textbf{Std Dev}\\
    \midrule
    OW & 0.0083 & 0.0149 & 0.0128\\
    2NN & 0.0225 & 0.0247 & 0.0016\\
    4NN & \textbf{0.0041} & \textbf{0.0046} & \textbf{0.0006}\\
    \bottomrule
  \end{tabular}
\end{table}

     

\begin{figure}
     \centering
     \begin{subfigure}{0.4\textwidth}
         \centering
         \includegraphics[width=\textwidth]{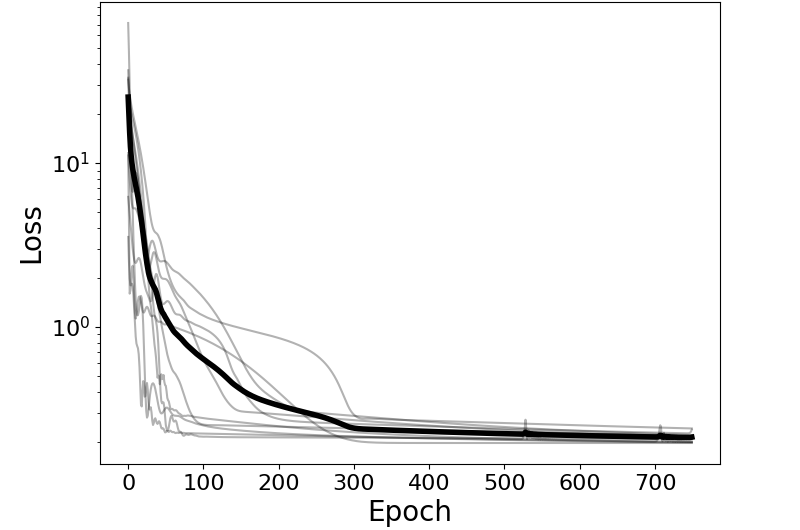}
         \caption{}
         \label{OWLossOW}
     \end{subfigure}
     \hspace{0.05\textwidth} 
     \begin{subfigure}{0.4\textwidth}
         \centering
         \includegraphics[width=\textwidth]{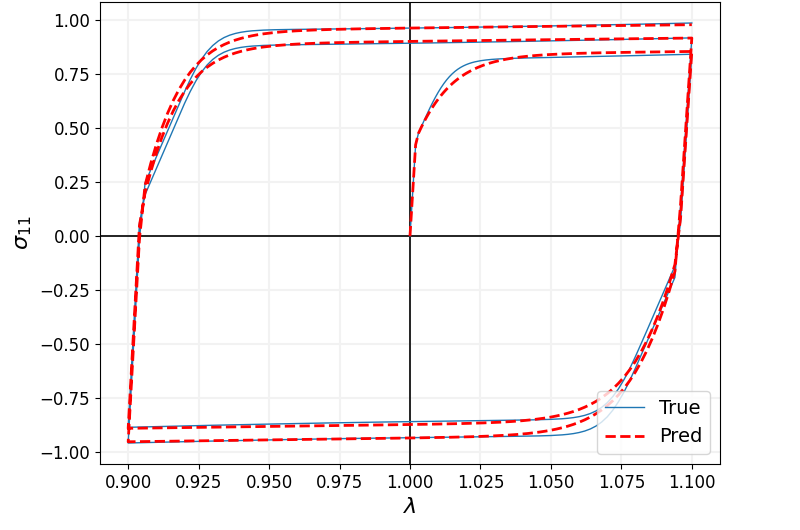}
         \caption{}
         \label{AFResponseOW}
     \end{subfigure}
     
     \caption{Results for AF on the second synthetic dataset. (a) Evolution of the loss during the training routine for fitting the parameters to Armstrong-Frederick model for kinematic hardening. The lighter lines represent the loss evolution for different initialization whereas the darker line represents the mean loss. (b) True and predicted stress values from the best loss response.}
     \label{AFDataOW}
\end{figure}

\begin{figure}
     \centering
     \begin{subfigure}{0.4\textwidth}
         \centering
         \includegraphics[width=\textwidth]{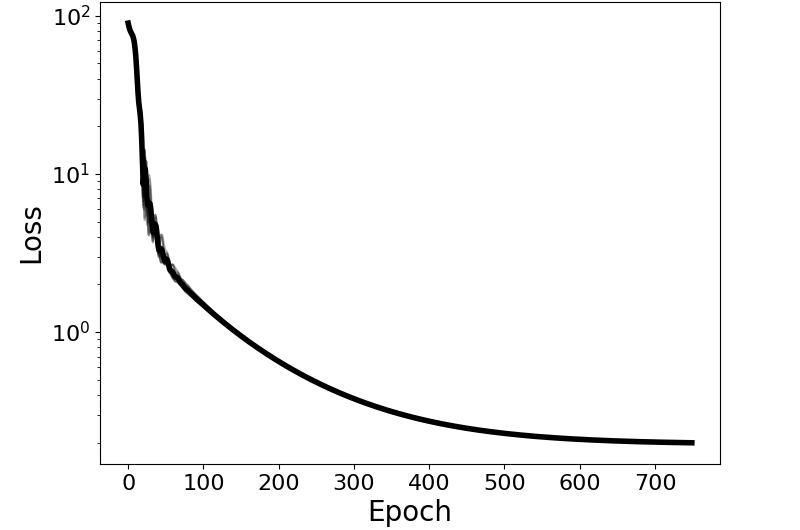}
         \caption{}
         \label{2NNLossOW}
     \end{subfigure}
     \hspace{0.05\textwidth} 
     \begin{subfigure}{0.4\textwidth}
         \centering
         \includegraphics[width=\textwidth]{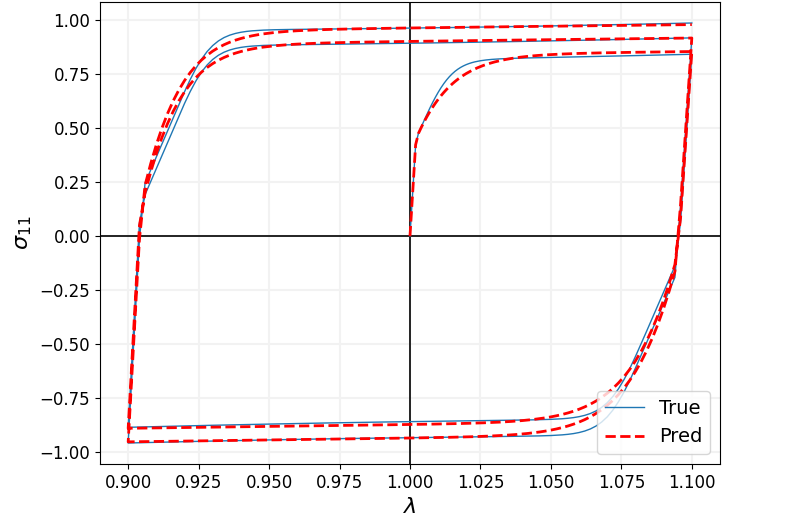}
         \caption{}
         \label{2NNResponseOW}
     \end{subfigure}
          \begin{subfigure}{0.4\textwidth}
         \centering
         \includegraphics[width=\textwidth]{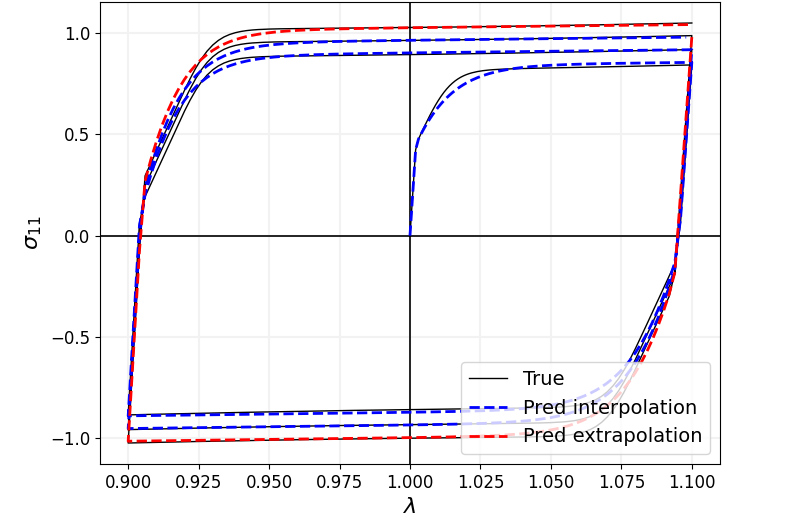}
         \caption{}
         \label{2NNExtrapolatedOW}
     \end{subfigure}
     
     \caption{Results for 2NN on the second synthetic dataset. (a) Evolution of the loss during the training routine for PANNs for dissipation potentials and fitting the parameters of Neo-Hookean free energies. The lighter lines represent the loss evolution for different initialization whereas the darker line represents the mean loss. (b) True and predicted stress values from the best loss response. (c) Model's performance on unseen data.}
     \label{2NNDataOW}
\end{figure}

\begin{figure}
     \centering
     \begin{subfigure}{0.4\textwidth}
         \centering
         \includegraphics[width=\textwidth]{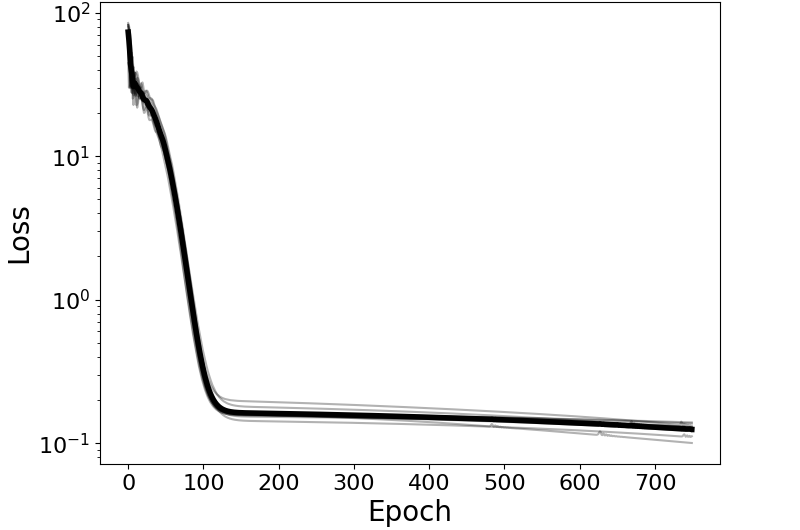}
         \caption{}
         \label{4NNLossOW}
     \end{subfigure}
     \hspace{0.05\textwidth} 
     \begin{subfigure}{0.4\textwidth}
         \centering
         \includegraphics[width=\textwidth]{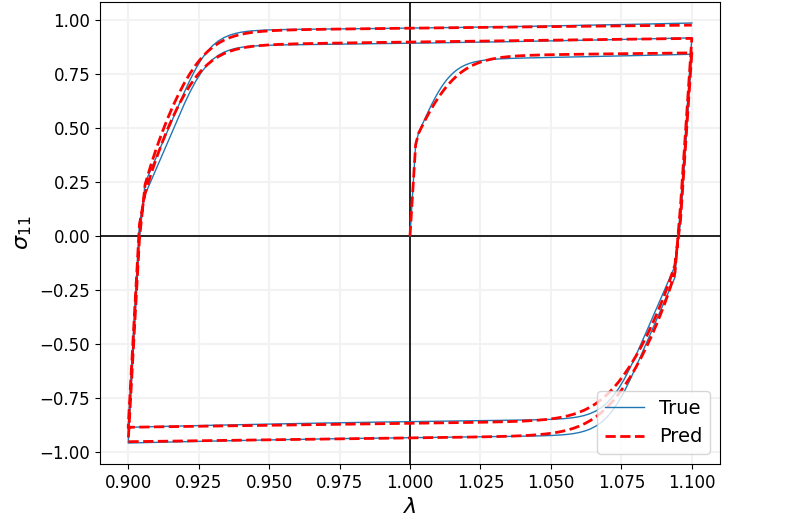}
         \caption{}
         \label{4NNResponseOW}
     \end{subfigure}
          \begin{subfigure}{0.4\textwidth}
         \centering
         \includegraphics[width=\textwidth]{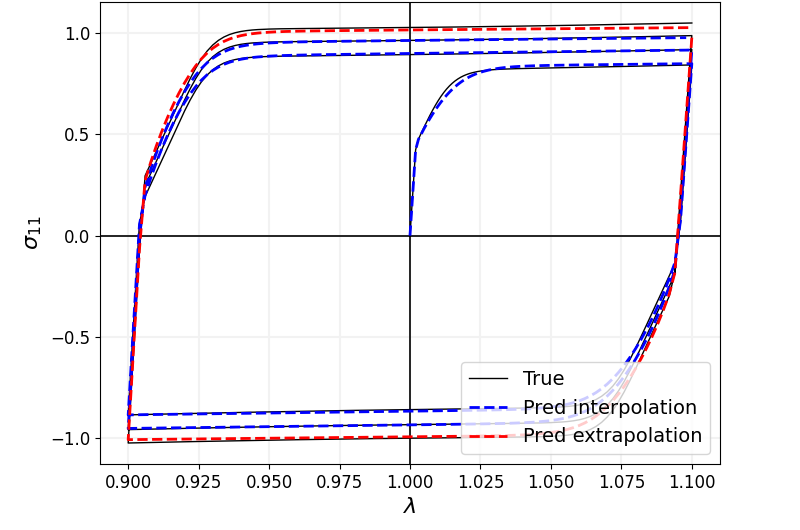}
         \caption{}
         \label{4NNExtrapolationOW}
     \end{subfigure}
     
     \caption{Results for 4NN on the second synthetic dataset. (a) Evolution of the loss during the training routine for PANNs for all the potentials. The lighter lines represent the loss evolution for different initialization whereas the darker line represents the mean loss. (b) True and predicted stress values from the best loss response. (c) Model's performance on unseen data.}
     \label{4NNDataOW}
\end{figure}

The second dataset was generated using the Ohno-Wang kinematic hardening model for the evolution of the kinematic dissipation potential and the parameters of the Armstrong-Frederick model were fitted as a baseline phenomenological law. Figure \ref{AFDataOW} shows the evolution of training losses and the best stress fits for this dataset. Figure \ref{2NNDataOW} depicts the same for the 2NN approach, where parameters of Neo-Hookean free energies were trained and PANNs were used for dissipation potentials. Figure \ref{4NNDataOW} highlights the loss evolution and best stress fit for the 4NN approach, utilizing PANNs for all energy and dissipation potentials. Extrapolation capabilities of the PANN-based approaches were also tested as shown in figures \ref{2NNExtrapolatedOW} and \ref{4NNExtrapolationOW}. The model was trained on 5 cycles of uniaxial loading and was then subjected to an additional two cycles for the model to extrapolate. Table \ref{tab:SynData_loss_table_OW} compares the approaches with 10 random initializations based on the lowest obtained loss, mean loss across all runs and standard deviation.

\begin{table}
  \centering
  \caption{Loss for each framework for synthetic data generated with the Ohno-Wang\cite{OHNO1993375, OHNO1993391} model.}
  \label{tab:SynData_loss_table_OW}
  \begin{tabular}{cccc}
    \toprule
    \textbf{Model} & \textbf{Lowest Loss} & \textbf{Mean Loss} & \textbf{Std Dev}\\
    \midrule
    AF & 0.1958 & 0.2047 & 0.0094\\
    2NN & 0.1959 & 0.1960 & \textbf{4.4e-5}\\
    4NN & \textbf{0.0902} & \textbf{0.1144} & 0.0117\\
    \bottomrule
  \end{tabular}
\end{table}

\subsection{Experimental Data}
Next, we tested the framework against experimental data obtained from cyclic uniaxial loading tests conducted on a mild steel sheet, as reported in \cite{yoshida2002elastic}. This dataset is particularly challenging due to the presence of transient Bauschinger deformation and work-hardening stagnation. The authors of \cite{yoshida2002elastic} and \cite{YOSHIDA2002661} introduced a new constitutive model, termed the Yoshida-Uemori model, to address the complexities in fitting this particular experimental data.

\begin{figure}
     \centering
     \begin{subfigure}{0.4\textwidth}
         \centering
         \includegraphics[width=\textwidth]{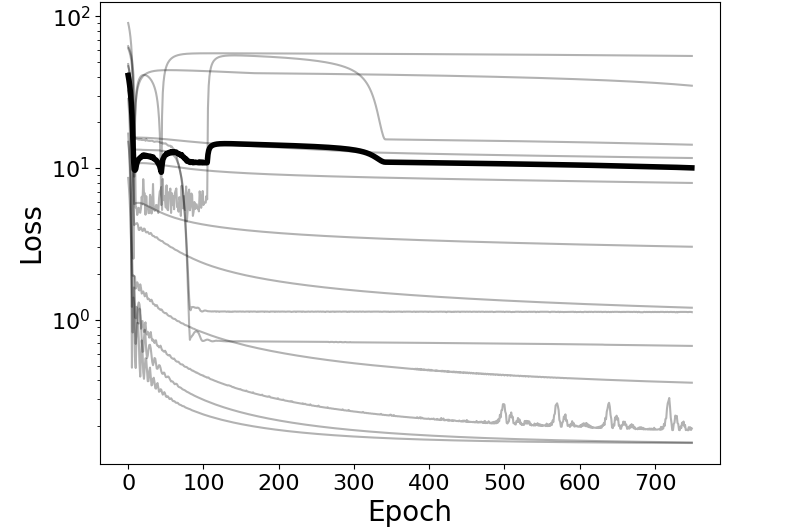}
         \caption{}
         \label{AFLossExp}
     \end{subfigure}
     \hspace{0.05\textwidth} 
     \begin{subfigure}{0.4\textwidth}
         \centering
         \includegraphics[width=\textwidth]{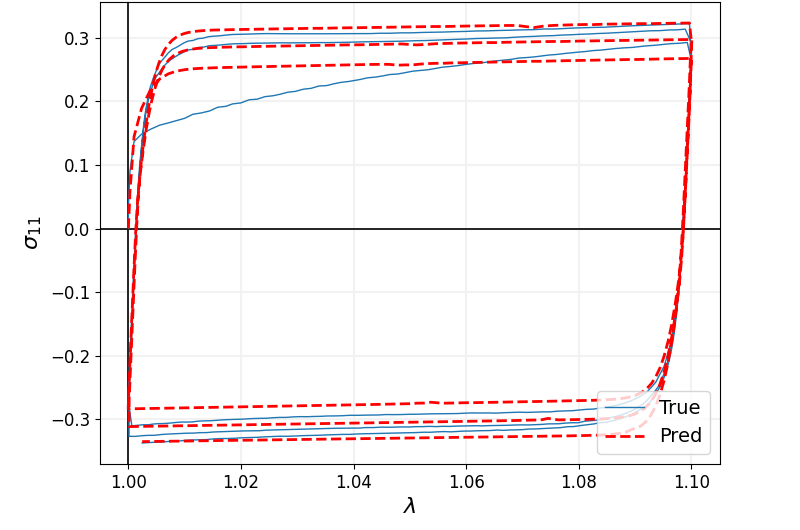}
         \caption{}
         \label{AFResponseExp}
     \end{subfigure}
     
     \caption{Results for AF on the experimental dataset. (a) Evolution of the loss during the training routine for fitting the parameters to Armstrong-Frederick model for kinematic hardening. The lighter lines represent the loss evolution for different initialization whereas the darker line represents the mean loss. (b) True and predicted stress values from the best loss response.}
     \label{AFDataExp}
\end{figure}

\begin{figure}
     \centering
     \begin{subfigure}{0.4\textwidth}
         \centering
         \includegraphics[width=\textwidth]{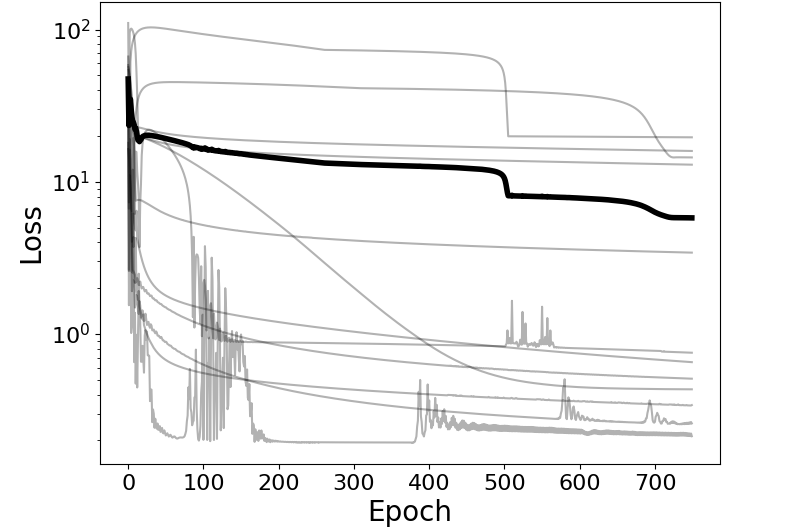}
         \caption{}
         \label{OWLossExp}
     \end{subfigure}
     \hspace{0.05\textwidth} 
     \begin{subfigure}{0.4\textwidth}
         \centering
         \includegraphics[width=\textwidth]{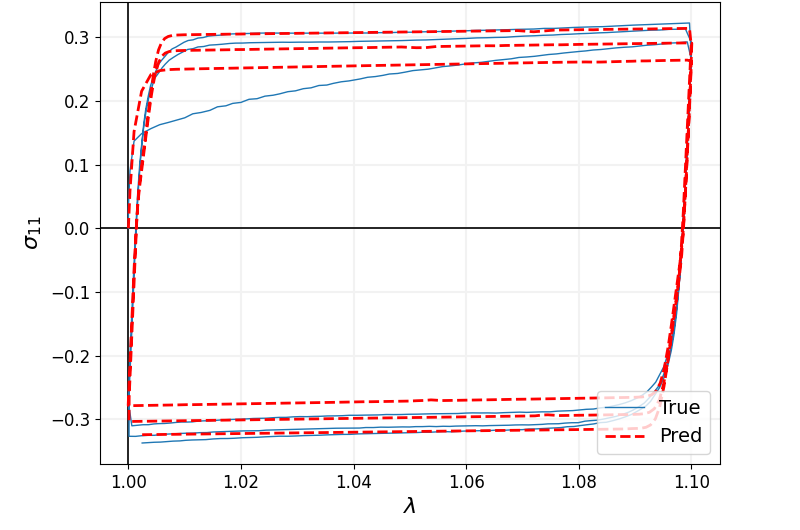}
         \caption{}
         \label{OWResponseExp}
     \end{subfigure}
     
     \caption{Results for OW on the experimental dataset. (a) Evolution of the loss during the training routine for fitting the parameters to Ohno-Wang model for kinematic hardening. The lighter lines represent the loss evolution for different initialization whereas the darker line represents the mean loss. (b) True and predicted stress values from the best loss response.}
     \label{OWDataExp}
\end{figure}

\begin{figure}
     \centering
     \begin{subfigure}{0.4\textwidth}
         \centering
         \includegraphics[width=\textwidth]{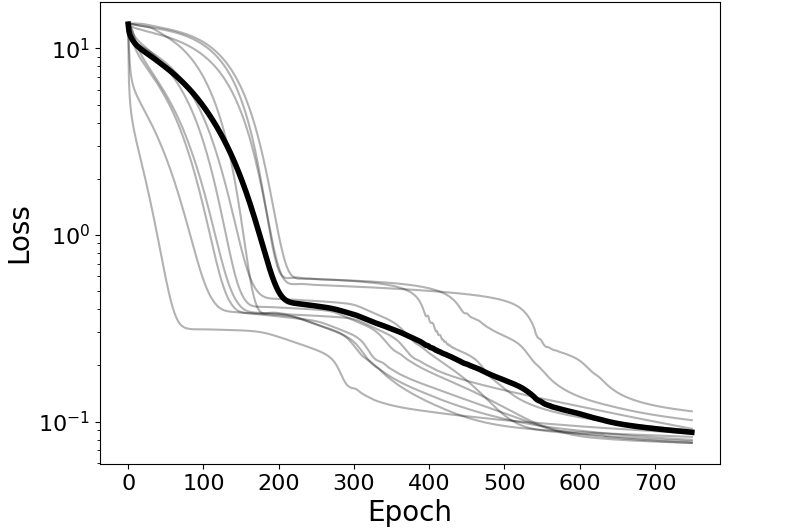}
         \caption{}
         \label{2NNLossExp}
     \end{subfigure}
     \hspace{0.05\textwidth} 
     \begin{subfigure}{0.4\textwidth}
         \centering
         \includegraphics[width=\textwidth]{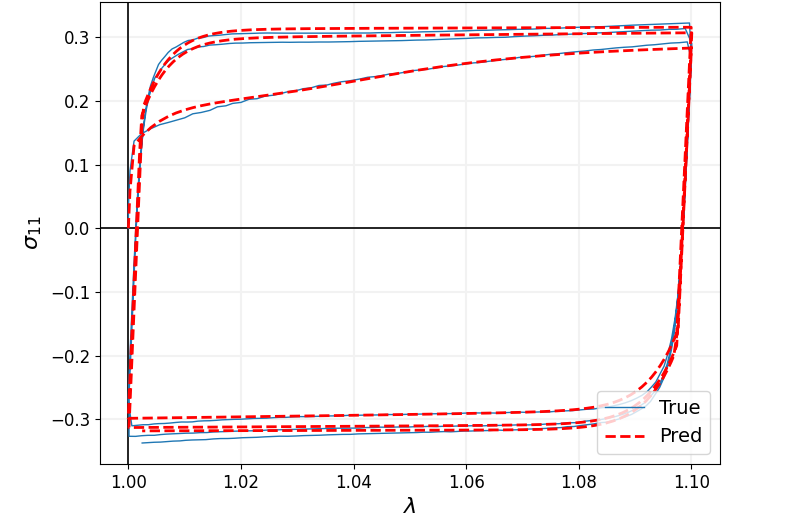}
         \caption{}
         \label{2NNResponseExp}
     \end{subfigure}
     
     \caption{Results for 2NN on the experimental dataset. (a) Evolution of the loss during the training routine for PANNs for dissipation potentials and fitting the parameters of Neo-Hookean free energies. The lighter lines represent the loss evolution for different initialization whereas the darker line represents the mean loss. (b) True and predicted stress values from the best loss response.}
     \label{2NNDataExp}
\end{figure}

\begin{figure}
     \centering
     \begin{subfigure}{0.4\textwidth}
         \centering
         \includegraphics[width=\textwidth]{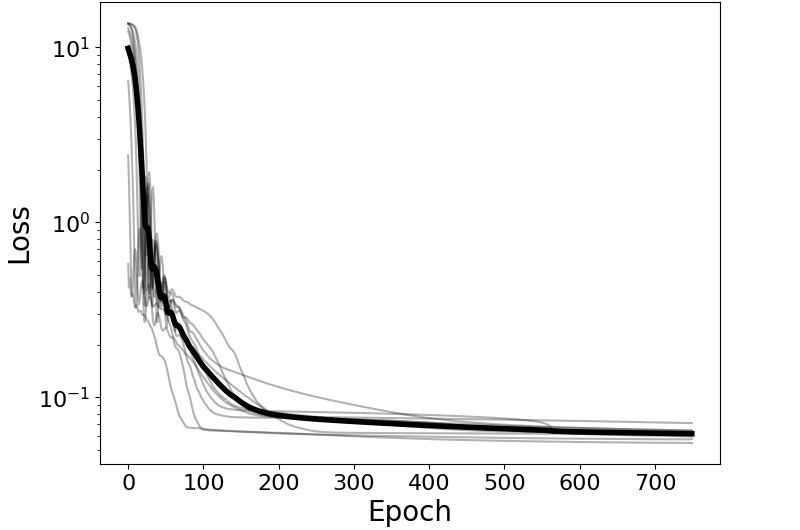}
         \caption{}
         \label{4NNLossExp}
     \end{subfigure}
     \hspace{0.05\textwidth} 
     \begin{subfigure}{0.4\textwidth}
         \centering
         \includegraphics[width=\textwidth]{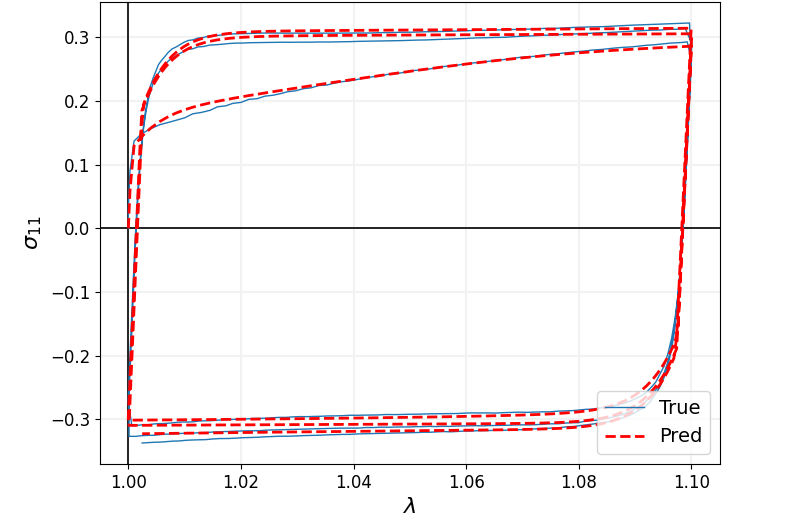}
         \caption{}
         \label{4NNResponseExp}
     \end{subfigure}
     
     \caption{Results for 4NN on the experimental dataset. (a) Evolution of the loss during the training routine for PANNs for all the potentials. The lighter lines represent the loss evolution for different initialization whereas the darker line represents the mean loss. (b) True and predicted stress values from the best loss response.}
     \label{4NNDataExp}
\end{figure}

Similar to synthetic data, the frameworks were subjected to 10 varying parameter initializations. Figures \ref{AFDataExp} and \ref{OWDataExp} present the evolution of training losses and the best stress fit for training the parameters of AF and OW kinematic hardening models, respectively. Figure \ref{2NNDataExp} showcases the results for the 2NN approach, while figure \ref{4NNDataExp} shows the loss evolution and best stress fit for the 4NN approach. Table \ref{tab:ExpData_loss_table} compares the lowest losses, mean losses, and the standard deviations obtained for each framework, highlighting their performance against the experimental data.

\begin{table}
  \centering
  \caption{Loss for each framework for experimental data.}
  \label{tab:ExpData_loss_table}
  \begin{tabular}{cccc}
    \toprule
    \textbf{Model} & \textbf{Lowest Loss} & \textbf{Mean Loss} & \textbf{Std Dev}\\
    \midrule
    AF & 0.1547 & 10.031 & 15.976\\
    OW & 0.2132 & 5.7917 & 7.2151\\
    2NN & 0.0689 & 0.0778 & 0.0077\\
    4NN & \textbf{0.0541} & \textbf{0.0598} & \textbf{0.0036}\\
    \bottomrule
  \end{tabular}
\end{table}

\subsection{Discussion}
 As highlighted in tables \ref{tab:SynData_loss_table_BC2} and \ref{tab:SynData_loss_table_OW}, the parameter-fitting approach works well on synthetic data, with comparable performance to the 2NN approach in terms of stress-fit while being inferior in terms of varying initial parameters. The 4NN approach shows the best performance across all metrics for synthetic data except for standard deviation for the Ohno-Wang data which is lowest for the 2NN framework. However, the 4NN framework outperforms the 2NN one in terms of stress response, and having a small standard deviation is of little value if the loss for stress fit is high. The parameter-fitting approaches show poor performance when tested on experimental data. Both the PANN-based frameworks show robustness against varying initializations whereas parameter-fitting approaches are prone to get stuck in local minima and are heavily reliant on the initial parameters. Clearly, the parameter-fitting approach has been applied successfully in the literature, typically utilizing gradient-free optimization algorithms combined with expert knowledge on hyperparameter tuning and distributions of multiple initial parameters. However, the PANN-based approaches give potentials that are able to fit the stress response, without requiring expert knowledge. Similar to the synthetic dataset, the 4NN approach shows the best performance among all with the 2NN approach giving comparable results. PANN-based frameworks have much more flexibility since the potentials, although constrained to some extent for thermodynamic consistency, can easily adapt to any form. Since the 4NN approach has four such potentials, it is expected to perform better than the 2NN approach where the free energies have fixed forms with only trainable scaling parameters. Since one can easily model these potentials using shallow neural networks, the difference between training two neural networks to training four is not significant enough to advocate the use of the 2NN framework in place of the 4NN one which shows exceptional performance on all datasets for both interpolation and extrapolation.

\section{Conclusions}
In this work, we introduce an NN-based framework designed to automate the discovery of constitutive models for finite strain elastoplasticity, using data from single uniaxial experiments. By employing physics-augmented NNs, we ensured that our framework is thermodynamically consistent by construction. The framework demonstrated reliable performance on both synthetic and experimental data, with particularly impressive extrapolation capabilities observed on synthetic datasets. This can be attributed to the integration of physical principles in the system which also allows us to discover constitutive models from a single uniaxial experiment whereas other NN-based frameworks that are ignorant of the physical constraints often require more data and perform poorly on unseen data. The comparison between NN-based frameworks and traditional parameter-fitting of phenomenological laws showed the robustness of proposed frameworks against different parameter initializations. We found that the fitting process for phenomenological laws is more susceptible to getting trapped in local minima, leading to convergence issues. High mean losses and standard deviations observed in models employing AF and OW kinematic hardening highlight the dependence on initial parameter values for achieving optimal results. While advanced optimization algorithms could improve the parameter search for phenomenological models, their usage typically requires expert knowledge whereas the objective of this study was to fully automate the model discovery process. Although the resulting models may lack interpretability, incorporating physics into our NNs yields a reliable framework that inherently satisfies physical, mechanical, and thermodynamic constraints. This automated discovery framework paves the way for future research on NN-based constitutive modeling of history-dependant materials in the finite strain regime. Additionally, it mitigates user bias and gives a more systematic, physics-informed framework that can easily be integrated into existing solvers. In the next steps, we are considering extending the framework with sparse regression techniques to make the obtained models more interpretable \cite{fuhg2024extreme} and adding temperature- and microstructure dependencies to the presented approach.

\clearpage
\appendix
\section*{Appendix A. Positivity of dissipation}
Recall the final form of dissipation potential from Equation \eqref{eq:finaldissipation}: 

\begin{align}
    \mathcal{D} &= \dot{\lambda} \left[ \left[\ts{M}-\ts{M}\kin\right]:\ts{\nu}  + \ts{M}\kin:\pdiff[\hat{\varPhi}\kin]{\ts{M}\kin} + \kappa \pdiff[\hat{\varPhi}\iso]{\kappa} \right], \quad \text{where} \quad \ts{\nu} := \pdiff[\varPhi]{\ts{M}}
\end{align}
The effective stress (Equation \eqref{eq:yield}), in our case von Mises, is a positive homogeneous function of degree one, such that
\begin{align}
    f\subscr{y}(\ts{M}\red) = \pdiff[f\subscr{y}]{\ts{M}\red} : \ts{M}\red
\end{align}
following Euler's homogeneous function theorem. Consequently, 
\begin{align}
    \ts{M}\red : \pdiff[\varPhi]{\ts{M}} = f\subscr{y}(\ts{M}\red) \geq 0
\end{align}

Convexity is defined as:

\begin{equation}
f(x_1) \geq f(x_2) + f'(x_2)(x_1 - x_2) \label{eq:convexity}
\end{equation}

Positivity of the term $\ts{M}\kin:\pdiff[\hat{\varPhi}\kin]{\ts{M}\kin}$ is ensured by making $\hat{\varPhi}\kin$ convex. Invoking the convexity definition from \eqref{eq:convexity}, we can write
\begin{equation}
    \hat{\varPhi}\kin (\ts{M}\kinn1) \geq \hat{\varPhi}\kin (\ts{M}\kinn2) + \pdiff[\hat{\varPhi}\kin]{\ts{M}\kinn2} : (\ts{M}\kinn1 - \ts{M}\kinn2)
\end{equation}
Setting $\ts{M}\kinn1=\ts{0}$, knowing $\hat{\varPhi}\kin(\ts{0})=0$, and $\ts{M}\kinn2= \ts{M}\kin$, i.e. any arbitrary tensor, we get
\begin{equation}
    0 \geq \hat{\varPhi}\kin (\ts{M}\kin) + \pdiff[\hat{\varPhi}\kin]{\ts{M}\kin} : -\ts{M}\kin
\end{equation}
\begin{equation}
    \pdiff[\hat{\varPhi}\kin]{\ts{M}\kin} : \ts{M}\kin \geq \hat{\varPhi}\kin (\ts{M}\kin)
\end{equation}
Since $\hat{\varPhi}\kin (\ts{M}\kin)$ is always non-negative by construction, $\pdiff[\hat{\varPhi}\kin]{\ts{M}\kin} : \ts{M}\kin$ will also always be non-negative.

The final term in \eqref{eq:finaldissipation} corresponding to isotropic hardening is given as as
\begin{equation}
    \kappa \frac{\partial\hat{\Phi}_{\text{iso}}}{\partial\kappa} \geq 0
\end{equation}
Since we only consider isotropic hardening, $\kappa$ is always greater or equal to zero, this implies
\begin{equation}
    \frac{\partial\hat{\Phi}_{\text{iso}}}{\partial\kappa} \geq 0
\end{equation}

which can simply be ensured my making $\hat{\Phi}_{\text{iso}}$ monotonically nondecreasing with respect to $\kappa$.

\section*{Appendix B. A note on the normalization of neural network outputs and derivatives}

   \begin{figure}[h]
    \centering
    \includegraphics[width=0.5\textwidth]{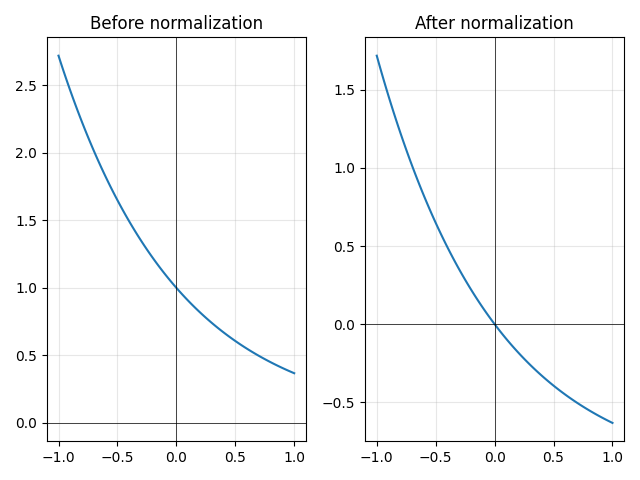}
    \caption{Violation of the positivity constraint upon normalization}
    \label{fig:normalizationprob}
\end{figure}

In data-driven constitutive modeling frameworks, the energy and dissipation potentials are typically defined using neural networks. Physically, we assume, that we are constrained to have zero energy and dissipation in the undeformed configuration. Furthermore, we enforce the stresses to be zero at the undeformed configuration, which are obtained as the derivatives of the free energies. That is, we require:
\[
\quad \hat{\varPhi}\kin (\ts{M}\kin = \ts{0}) = 0 \quad \text{and} \quad  \hat\varPhi\iso({\kappa} = 0) = 0
\]
\[
\quad \hat{\varPsi}\kin (\ts{c}\kin = \mathbf{I}) = 0 \quad \text{and} \quad \pdiff[\hat{\varPsi}\kin]{\ts{c}\kin} \bigg|_{\ts{c}\kin = \mathbf{I}} = 0
\]
\[
\quad \Psi_{iso}({k} = 0) = 0 \quad \text{and} \quad \frac{\partial \Psi_{iso}}{\partial {k}} \bigg|_{{k} = 0} = 0
\]
While it is straightforward to ensure this by merely translating the neural network outputs or its derivatives, this translation might lead to violations of the imposed constraints. Consider, for example, that we require a positive and convex neural network. Upon normalization, we may violate the positivity constraint as shown in Figure \ref{fig:normalizationprob}. Similarly, when normalizing the derivatives to have a stress-free reference state, we need to be careful about the physical constraints. Consider a convex, monotonically increasing neural network with $f$ representing the output of the network for an input $x$:
\[
f''(x) \geq 0 \quad \text{and} \quad f'(x) \geq 0.
\]

The condition \( f''(x) \geq 0 \) implies that \( f'(x) \) is monotonically increasing. This is because the first derivative \( f'(x) \) is itself a function whose derivative \( f''(x) \) is non-negative, indicating that \( f'(x) \) does not decrease. So when we normalize this derivative, we violate the monotonically increasing constraint for our NN since normalization will lead to:
\[
\quad f'_{\text{norm}}(x) < 0 \quad \text{for} \quad x < 0
\]
Similarly, the derivative of a convex, monotonically decreasing function would also be monotonically increasing and the translation of these derivatives to ensure they are zero at the undeformed configuration violates the monotonically decreasing constraint in the positive domain.

\section*{Appendix C. Fitting model parameters to itself}
\begin{figure}[h]
     \centering
     \begin{subfigure}{0.4\textwidth}
         \centering
         \includegraphics[width=\textwidth]{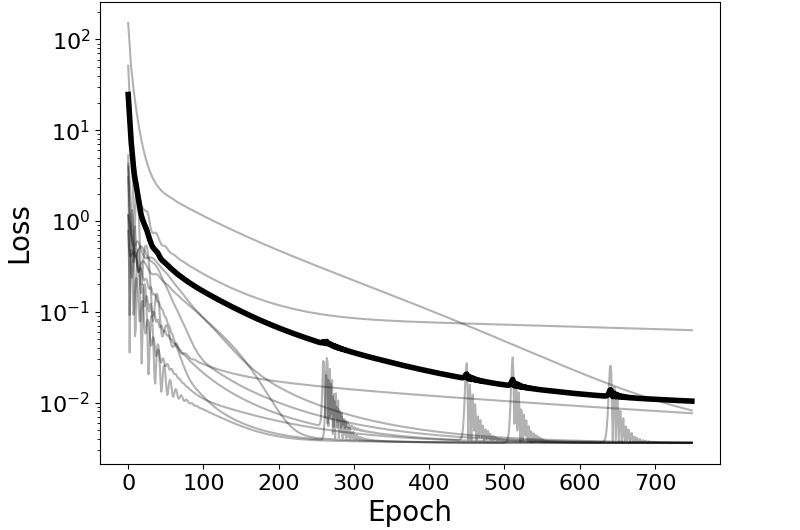}
         \caption{}
         \label{AFLossBC2}
     \end{subfigure}
     \hspace{0.05\textwidth} 
     \begin{subfigure}{0.4\textwidth}
         \centering
         \includegraphics[width=\textwidth]{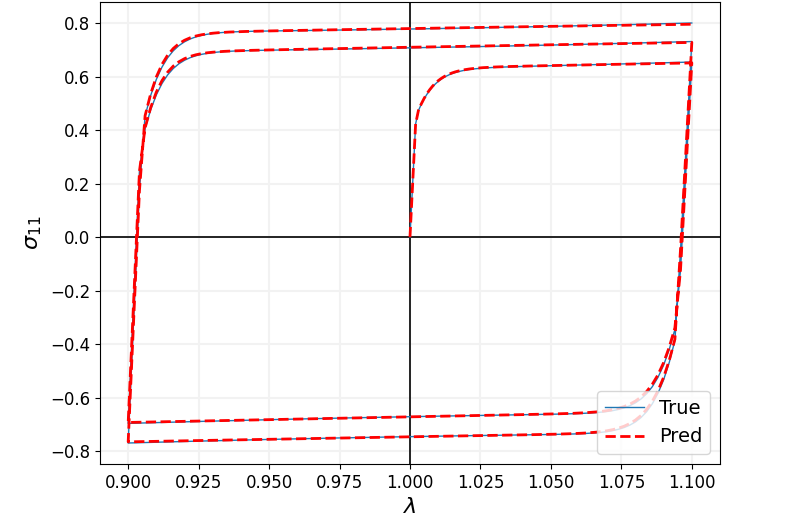}
         \caption{}
         \label{AFResponseBC2}
     \end{subfigure}
     
     \caption{(a) Evolution of the loss during the training routine for fitting the parameters to Armstrong-Frederick model for kinematic hardening. The lighter lines represent the loss evolution for different initialization whereas the darker line represents the mean loss. (b) True and predicted stress values from the best loss response. Here, $\lambda$ represents the stretch and the subscript $11$ represents the component of Cauchy stress $\sigma$ (GPa) along the direction of uniaxial compression or tension.}
     \label{AFDataBC2}
\end{figure}

\begin{figure}[h]
     \centering
     \begin{subfigure}{0.4\textwidth}
         \centering
         \includegraphics[width=\textwidth]{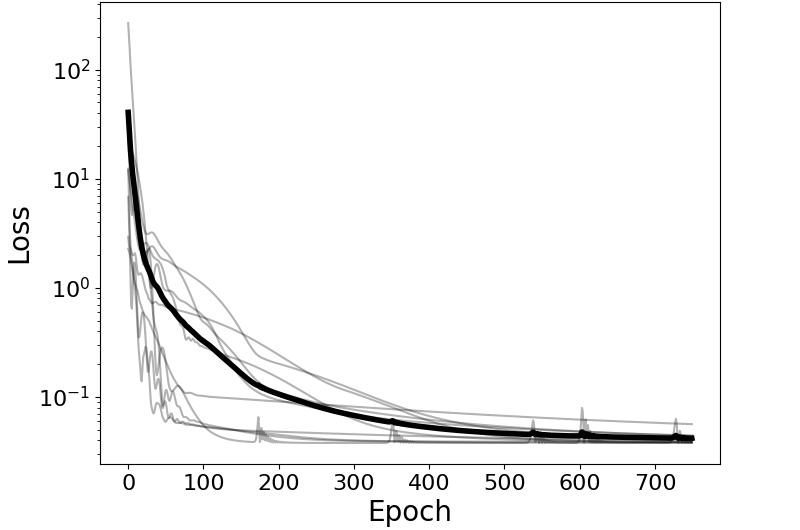}
         \caption{}
         \label{OWLossOW}
     \end{subfigure}
     \hspace{0.05\textwidth} 
     \begin{subfigure}{0.4\textwidth}
         \centering
         \includegraphics[width=\textwidth]{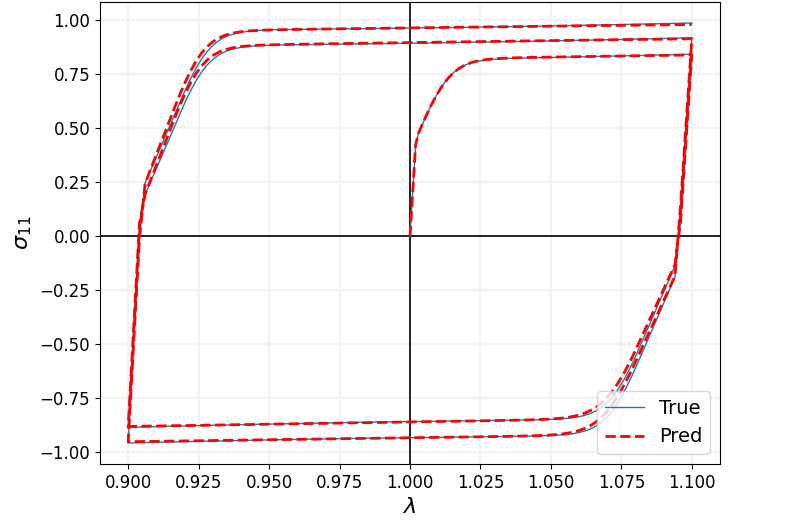}
         \caption{}
         \label{OWResponseOW}
     \end{subfigure}
     
     \caption{(a) Evolution of the loss during the training routine for fitting the parameters to Armstrong-Frederick model for kinematic hardening. The lighter lines represent the loss evolution for different initialization whereas the darker line represents the mean loss. (b) True and predicted stress values from the best loss response. Here, $\lambda$ represents the stretch and the subscript $11$ represents the component of Cauchy stress $\sigma$ (GPa) along the direction of uniaxial compression or tension.}
     \label{OWDataOW}
\end{figure}

While presenting the results, we omitted the parameter-fitting training for the model used to generate the data since it is a trivial task and we were mainly interested in examining the stress responses coming from training different models than the ones used for generating the dataset. However, the results for training the model parameters on the model itself are presented here. Figures \ref{AFDataBC2} and \ref{OWDataOW} show the evolution of loss and best stress fit for the parameter-fitting process of Armstrong-Frederick and Ohno-Wang kinematic hardening models for both the synthetic datasets respectively. It can be seen that the model is able to fit itself well.

\clearpage
\bibliographystyle{unsrt}
\bibliography{references.bib}

\end{document}